\documentclass[useAMS,usenatbib]{mn2e}
\usepackage[dvips]{epsfig}
\usepackage[]{natbib}
\usepackage{times}
\usepackage[]{fixltx2e}
\title[Gas Inflows towards the nucleus of NGC\,7213]{Gas inflows towards the nucleus of the active galaxy NGC\,7213}
\author[A. Schnorr-M\"uller et al.]
  {Allan Schnorr-M\"uller,$^1$ Thaisa Storchi-Bergmann,$^1$ Neil M. Nagar,$^2$ Fabricio Ferrari$^3$\\
  $^1$Instituto de F\'isica, Universidade Federal do Rio Grande do Sul, 91501-970, Porto Alegre, RS, Brazil\\
  $^2$Astronomy Department, Universidad de Concepci\'on, Casilla 160-C, Concepci\'on, Chile\\
  $^3$Instituto de Matem\'atica, Estat\'istica e F\'isica, Universidade Federal do Rio Grande (FURG), 96201-900, Rio Grande, RS, Brazil\\}

\pagerange{\pageref{firstpage}--\pageref{lastpage}} 

\begin{document}
\label{firstpage}
\maketitle

\begin{abstract}
We present two-dimensional stellar and gaseous kinematics of the inner 0.8\,$\times$\,1.1\,kpc$^2$ of the LINER/Seyfert\,1 galaxy NGC\,7213, from optical spectra obtained with the GMOS integral field spectrograph on the Gemini South telescope at a spatial resolution of $\approx$\,60\,pc. The stellar kinematics shows an average velocity dispersion of 177\,km\,s$^{-1}$, circular rotation with a projected velocity amplitude of 50\,km\,s$^{-1}$ and a kinematic major axis at a position angle of $\approx$\,-4$^\circ$ (west of north). From the average velocity dispersion we estimate a black hole mass of M$_{BH}$\,=\,8\,$_{-6}^{+16}\times$10$^{7}$\,M$_{\odot}$. The gas kinematics is dominated by non-circular motions, mainly along two spiral arms extending from the nucleus out to $\approx$\,4\arcsec (280\,pc) to the NW and SE, that are cospatial with a nuclear dusty spiral  seen in a structure map of the nuclear region of the galaxy. The projected gas velocities along the spiral arms                                                                                      
show blueshifts in the far side and redshifts in the near side, with values of up to 200\,km\,s$^{-1}$. This kinematics can be interpreted as gas inflows towards the nucleus along the spiral arms if the gas is in the plane of the galaxy. We estimate the mass inflow rate  using two different methods. The first is based of the observed velocities and geometry of the flow, and gives a mass inflow rate in the ionised gas of $\approx$\,7\,$\times$\,10$^{-2}$\,M$_{\odot}$\,yr$^{-1}$. In the second method, we calculate  the   net ionised gas mass flow rate through concentric circles of decreasing radii around the nucleus resulting in mass inflow rates ranging from $\approx$\,0.4\,M$_{\odot}$\,yr$^{-1}$ at 300\,pc down to $\approx$\,0.2\,M$_{\odot}$\,yr$^{- 1}$ at 100\,pc from the nucleus. These rates are larger than necessary to power the active nucleus.
\end{abstract}

\section{Introduction}

In the last few years, our research group AGNIFS (AGN Integral Field Spectroscopy) has been mapping the gas kinematics of the inner kiloparsec of nearby active galaxies in search of signatures of gas inflows to the nucleus. This work was motivated by the finding of \citet{lopes07} that there is a marked difference in the dust and gas content of this regions in early-type active galaxies when compared to non-active ones: while the first always have dusty structures, in the form of spiral and filaments at hundred of parsec scales, only 25\% of the non-active ones have such structures. This indicates that a reservoir of gas and dust is a necessary condition for the nuclear activity and suggests that the dusty structures are tracers of feeding channels to the Active Galactic Nuclei (hereafter AGN).

Previous studies by our group using integral field spectroscopy of the inner kiloparsec in nearby active galaxies  in the optical have revealed inflows in ionised gas in NGC\,1097 \citep{fathi06}, NGC\,6951 \citep{thaisa07} and M\,81 \citep{allan11}. In the particular case of M\,81, we could obtain not only the gas kinematics, but also the stellar kinematics, which was compared to that of the gas in order to isolate non-circular motions, instead of relying solely on the modelling of the gaseous kinematics as we did for NGC\,1097 and NGC\,6951. In the near-infrared, our group observed inflows in the central few hundred of parsecs of  the galaxies NGC\,4051 \citep{rogemar08}, NGC\,4151 \citep{thaisa10}, Mrk\,1066 \citep{rogemar11a}, Mrk\,1157 \citep{rogemar11b} and Mrk\,79 \citep{rogemar13}.

In this work, we present a new case of inflows observed in the inner kiloparsec of a nearby active galaxy: NGC\,7213, a Sa galaxy harbouring a Seyfert 1 AGN, using integral field spectroscopic observations in the optical. NGC\,7213 is located at a distance of 23.6\,Mpc (from NED\footnote{NASA/IPAC extragalactic database}), corresponding to a scale of 115\,pc\,arcsec$^{-1}$. Its nucleus was classified as Seyfert 1 by \citet{phillips79} and later included in the LINER class by \citet{filippenko84}. The optical continuum image of the galaxy is dominated by an almost featureless bulge with no apparent sign of a recent interaction \citep{hameed99}. Observations of neutral and ionised gas, however, reveal a different picture: H$\alpha$ observations show a ring of H\,II regions at $\approx$\,2\,kpc from  the nucleus \citep{thaisa96} and a filament, with no counterpart in optical continuum images, 18.6\,kpc south of the nucleus. In addition, H\,I observations show that the filament is 
part of a larger H\,I tail, and that the overall morphology and kinematics in H\,I is highly disturbed, indicating the galaxy has undergone a recent merging event \citep{hameed01}.  

The present paper is organised as follows. In Section \ref{Observations} we describe the observations and data reduction. In Section \ref{Results} we present the procedures used for the analysis of the data and the subsequent results. In section \ref{Discussion} we discuss our results and present estimates of the mass inflow rate using two distinct methods and in Section \ref{Conclusion} we present our conclusions.

\begin{figure*}
\includegraphics[scale=0.8]{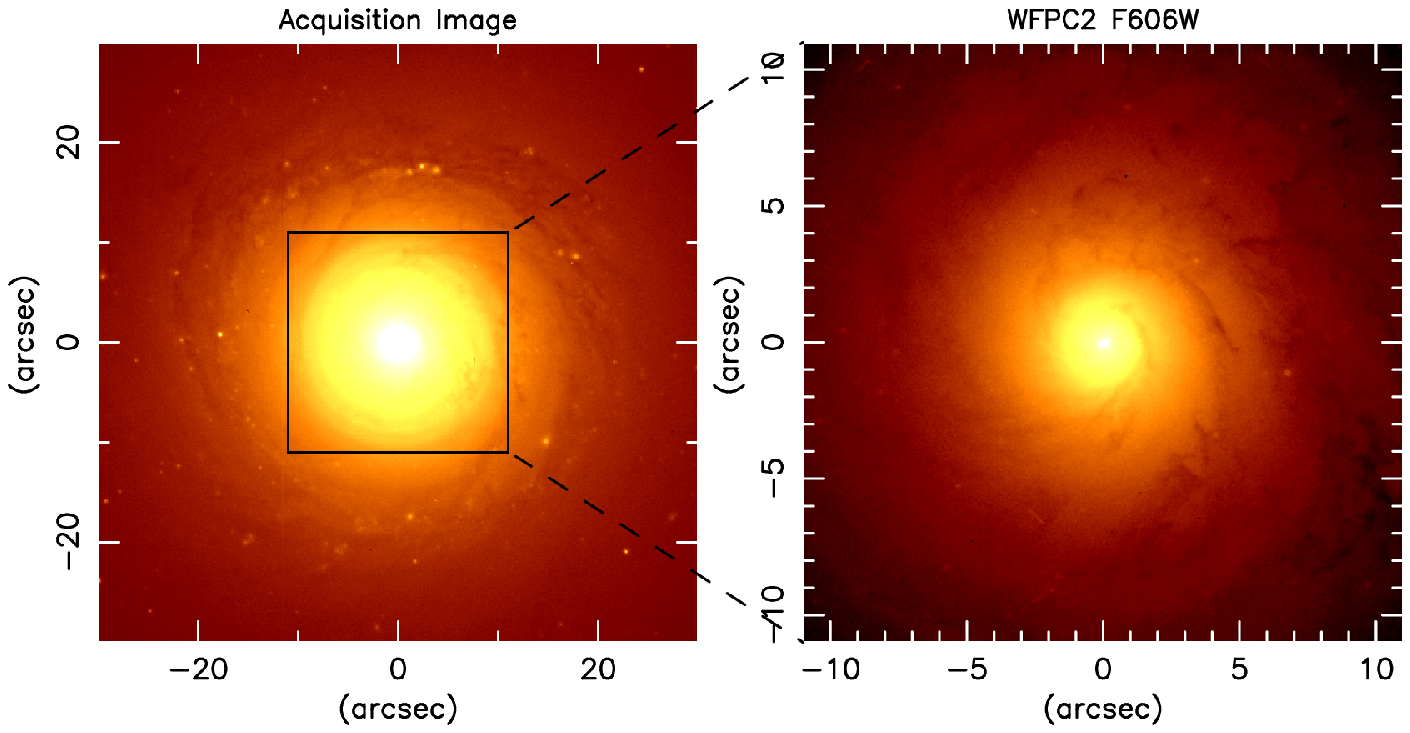}
\includegraphics[scale=0.8]{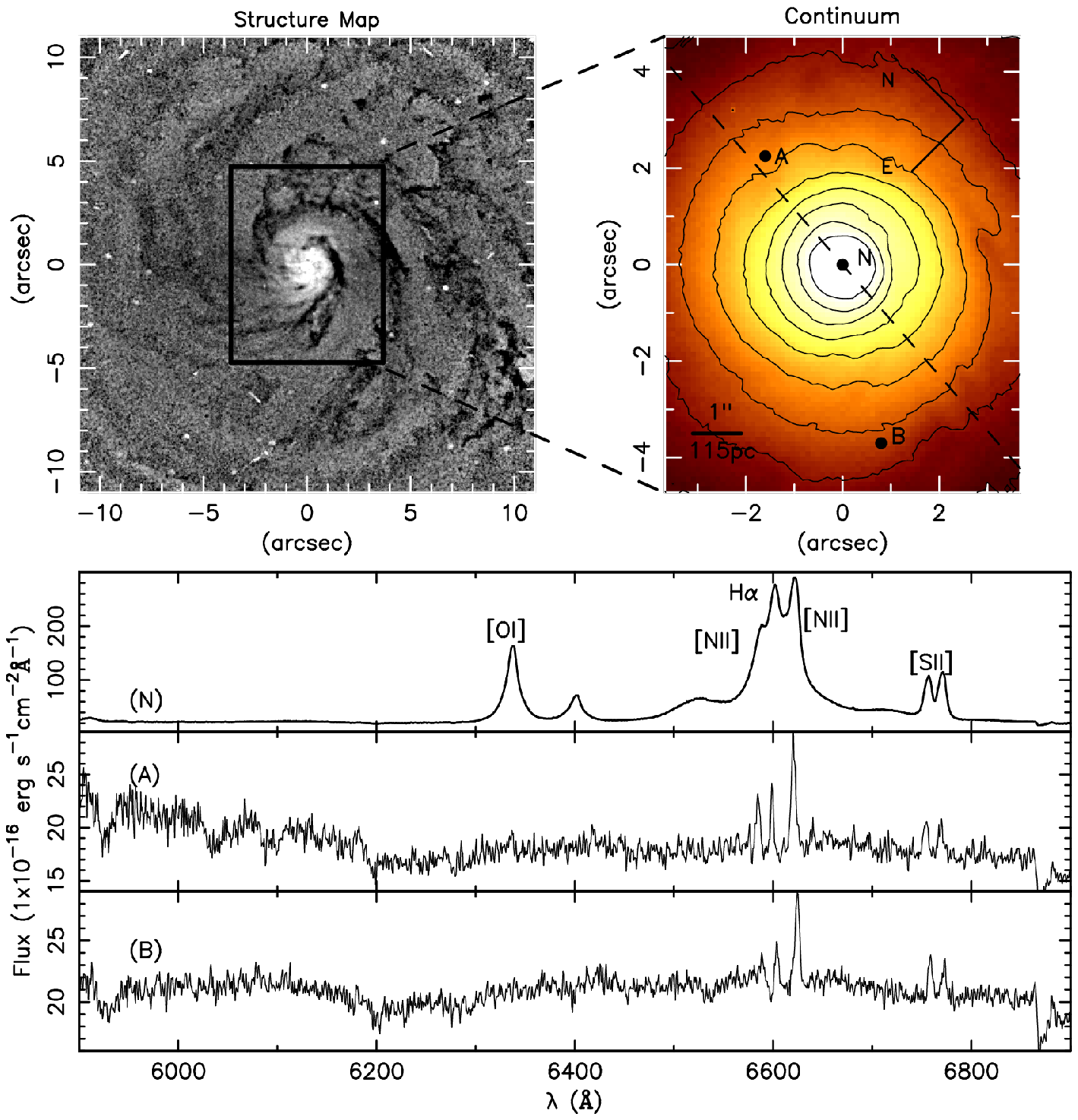}
\caption[Large scale image of NGC7213]{Top left: acquisition image of NGC\,7213. Top right: WFPC2 image of the nuclear region. Middle left: structure map. The rectangle shows the field-of-view of the IFU observations. Middle right: continuum image from the IFU spectra. The dashed white line indicates the position of the kinematic major axis of the galaxy (PA\,=\,-4\ensuremath{^\circ}). Bottom: spectra corresponding to the regions identified as N, A and B in the IFU image.}
\label{figure1}
\end{figure*}

\section {Observations and Data Reduction}\label{Observations}
The observations were obtained with the Integral Field Unit of the Gemini Multi Object Spectrograph (GMOS-IFU) at the Gemini South telescope on the night of September 27, 2011 (Gemini project GS-2011B-Q-23). The observations consisted of two adjacent IFU fields (covering 7\,$\times$\,5\,arcsec$^{2}$ each) resulting in a total angular coverage of 7\,$\times$\,10\,arcsec$^{2}$ around the nucleus. Six exposures of 350 seconds were obtained for each field, slightly shifted and dithered in order to correct for detector defects after combination of the frames. The seeing during the observation was 0\farcs5, as measured from the FWHM of a spatial profile of the calibration standard star. This corresponds to a spatial resolution at the galaxy of 58\,pc.

The selected wavelength range was 5600-7000\,\r{A}, in order to cover the H$\alpha$+[N\,II]\,$\lambda\lambda$6548,6583\,\r{A} and [S\,II]\,$\lambda\lambda$6716,6731\,\r{A} emission lines, observed with the grating GMOS R400-G5325 (set to central wavelength of either $\lambda$\,6500\,\r{A} or $\lambda$\,6550\,\r{A}) at a spectral resolution of R\,$\approx$\,2000.

The data reduction was performed using specific tasks developed for GMOS data in the
\textsc{gemini.gmos} package as well as generic tasks in \textsc{iraf}\footnote{\textit{IRAF} is distributed by the National Optical Astronomy Observatories, which are operated by the Association of Universities for Research in Astronomy, Inc., under cooperative agreement with the National Science Foundation.}. The reduction process comprised bias subtraction, flat-fielding, trimming, wavelength calibration, sky subtraction, relative flux calibration, building of the data cubes at a sampling of 0\farcs1$\,\times\,$0\farcs1, and finally the alignment and combination of the 12 data cubes.

\begin{figure*}
\includegraphics[scale=0.6]{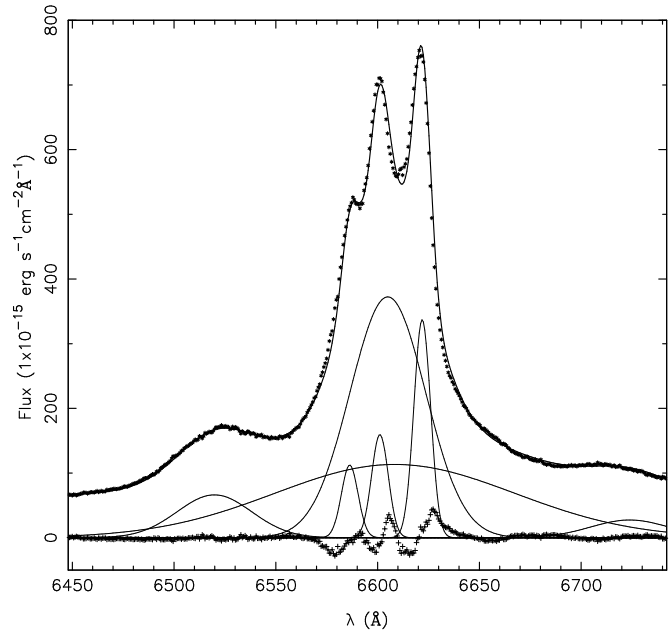}
\caption[Gaseous Kinematics]{Sample spectrum of the nucleus (asterisks) showing the individual Gaussians (thin solid lines) and the total fit (thick solid lines), as well as the residuals of the fit.}
\label{figure2}
\end{figure*}
\section{Results}\label{Results}

In Fig.\,\ref{figure1} we present, in the upper left panel, the acquisition image of NGC\,7213, where a ring of star-forming regions can be observed surrounding the nucleus at a radius $\approx$\,20\arcsec\,(thus beyond the field-of-view -- hereafter FOV -- of our observations). In the upper right panel we present an image of the inner 22\arcsec\,$\times\,$22\arcsec\ of the galaxy obtained with the WFPC2 (Wide Field Planetary Camera 2) through the filter F606W aboard the Hubble Space Telescope (hereafter HST). In the middle left panel we present a structure map of this image (see \citet{lopes07}), where the presence of nuclear dusty spirals is revealed. The rectangle shows the FOV covered by the IFU observations. In the middle right panel we present a continuum image obtained from our IFU observations by integrating the flux within a spectral window from $\lambda$6470\,\r{A} to $\lambda$6580\,\r{A}. The dashed line indicates the position of the kinematic major axis of the galaxy (position angle PA=-4\ensuremath{^\circ}), obtained from our measurements and model fit of the stellar kinematics (see section\,\ref{gaskinexc}). In the lower panel we present three spectra from the locations marked as N (nucleus), A and B, in the IFU image and extracted within apertures of 0\farcs3\,$\times\,$0\farcs3.

The nuclear spectrum (identified by N in Fig.\,\ref{figure1}) shows a broad, double-peaked H$\alpha$ component, which has led to the classification of NGC\,7213 as a Seyfert 1 galaxy, and also narrow [O\,I]\,$\lambda$$\lambda$\,6300,6363\,\r{A}, [N\,II]\,$\lambda$$\lambda$6548,6583\,\r{A}, H$\alpha$ and [S\,II]\,$\lambda$$\lambda$6717,6731\,\r{A} emission lines. The spectra from locations A and B shows only fainter emission in the [N\,II], H$\alpha$ and [S\,II] emission lines. In these extranuclear spectra, many absorption features from the stellar population are also clearly seen.

We adopt as the nucleus of the galaxy the position of the peak flux of the continuum, which corresponds also to the centroid of the emission in the broad double-peaked line.

\begin{figure*}
\includegraphics[scale=1.2]{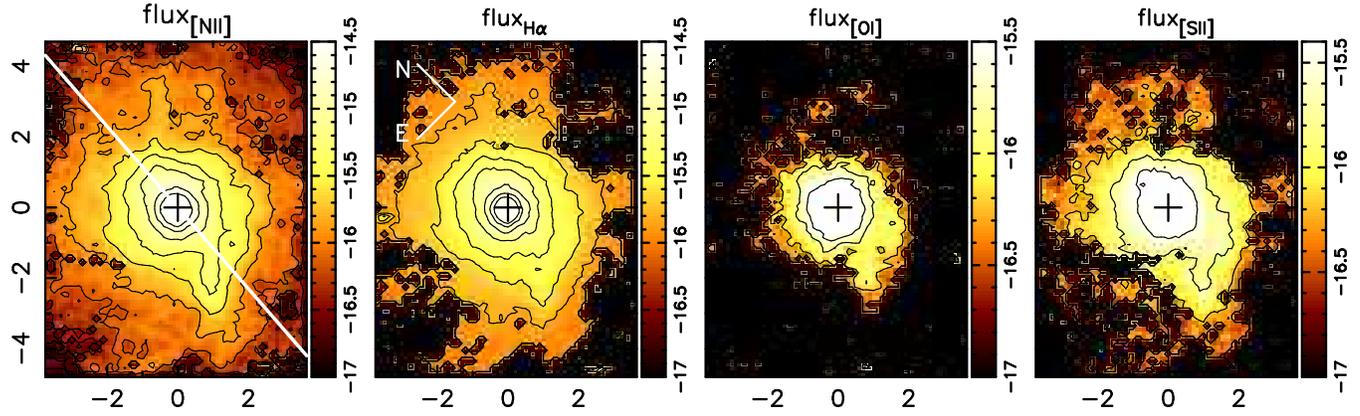}
\caption[Flux maps]{Maps of the [N\,II], H$\alpha$, [O\,I] and [S\,II] integrated fluxes in logarithmic scale (erg\,cm$^{-2}$\,s$^{-1}$ per pixel). }
\label{figure3}
\end{figure*}

\begin{figure*}
\begin{center}
\includegraphics[scale=1.2]{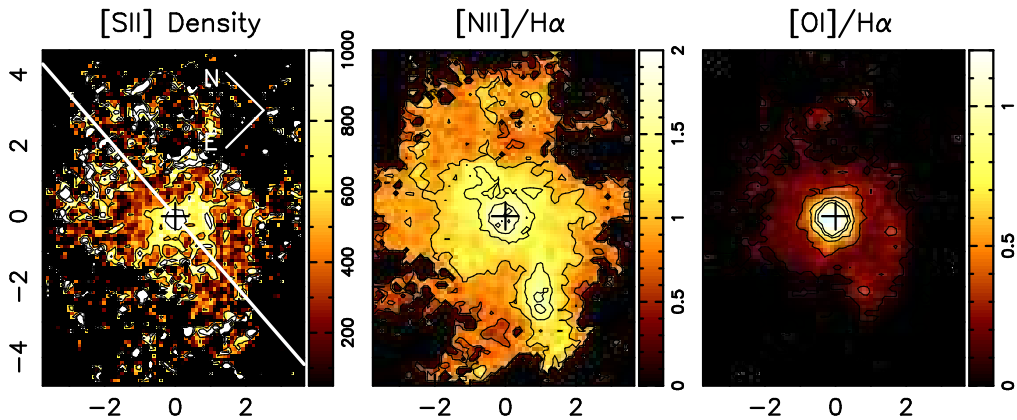}
\caption[Line ratio maps]{Gas density (cm$^{-3}$) distribution, [N\,II]/H$\alpha$ and [O\,I]/H$\alpha$ line ratio maps.}
\label{figure4}
\end{center}
\end{figure*}

\subsection{Measurements}

The gaseous centroid velocities, velocity dispersions and emission-line fluxes were obtained through the fit of Gaussians to the [N\,II], H$\alpha$ (narrow component), [O\,I] and [S\,II] emission lines. In our fit we adopted the following physically motivated constraints: 
\begin{enumerate}
\item Flux$_{[N\,II]\,\lambda6583}$/Flux$_{[N\,II]\,\lambda6548}=3$;
\item The [N\,II]\,$\lambda$6583 and [N\,II]\,$\lambda$6548 lines have the same centroid velocity and FWHM;
\item The narrow H$\alpha$ component and the [N\,II] lines have the same centroid velocity.
\end{enumerate}
One can argue that the H$\alpha$ and [N\,II] centroid velocity may not be the same, so we tested our assumption: in fits in which the H$\alpha$ centroid velocity was a free parameter, the difference between the H$\alpha$ and [N\,II] centroid velocities was always smaller than the error in the measurements. Considering this, we adopted the assumption (iii) above as it allowed a better measurement of the FWHM and flux at locations where H$\alpha$ was faint (having lower signal-to-noise ratio).

The flux distribution of the H$\alpha$ line was corrected for the contribution of underlying stellar absorption assuming an equivalent width of the stellar H$\alpha$ absorption line of 1.75\,\r{A}. This value was obtained measuring the equivalent width of the H$\alpha$ absorption in old stellar population models from \citet{bruzual03}, convolved with a Gaussian with $\sigma$\,=\,180\,km\,s$^{-1}$, a typical value of the stellar velocity dispersion in the centre of this galaxy (see section\,\ref{stellarkinematics}).

As the Point-Spread Function (PSF) has a width of 0\farcs5, within a radius of this order, the broad H$\alpha$ component is also present in the spectra. In order to isolate the narrow emission line, we have fitted and subtracted this broad component. In order to do that, four Gaussians were necessary: a ``blue'' and a ``red'' Gaussian to fit the blue and red peaks of the profiles, and another  two to fit the central part of the profile. The fit is illustrated in Fig.\,\ref{figure2}. The total flux of the central broad component is 8.9\,$\times$\,10$^{-12}$\,erg\,cm$^2$\,s$^{-1}$. Errors in all the measurements were estimated from Monte Carlo simulations in which Gaussian noise was added to the spectra; one hundred iterations were performed. 

In order to measure the stellar kinematics, we employed the Penalized Pixel Fitting technique (pPXF) \citep{cappellari04}. The \citet{bruzual03} models were used as template spectra. These models have a spectral  resolution ($\sigma$) of 61\,km\,s$^{-1}$ at 6300\AA, very similar to our value of 66\,km\,s$^{-1}$, so no corrections were made. Monte Carlo simulations based on the best-fitting absorption spectra obtained from the Penalized Pixel Fitting technique were also carried out to estimate the errors in the kinematic parameters. 

\subsection{Line fluxes and excitation of the emitting gas}\label{fluxgas}

In Fig.\ref{figure3} we present the integrated flux distributions in the [N\,II]\,$\lambda$6583\,\r{A}, H$\alpha$, [O\,I]\,$\lambda$6300\,\r{A} and [S\,II]\,$\lambda$6716\,\r{A} emission lines. 

The flux distributions show the highest values within the inner 2\arcsec(230\,pc) around the nucleus and extend  somewhat  farther from the nucleus (4\arcsec--460\,pc) to the SE, following one of the nuclear spiral arms seen in the structure map shown in Fig.\,\ref{figure1}. There is also an elongation observed to the north, at the beginning of the other spiral arm.

In Fig.\ref{figure4} we present the [N\,II]/H$\alpha$ and [O\,I]/H$\alpha$ line ratios and the gas density map, obtained from the [SII]\,$\lambda\lambda$6717/6731\,\r{A} line ratio assuming an electronic temperature of 10000K \citep{osterbrock89}. 

The gas density reaches a peak value of 1000\,cm$^{-3}$ at the nucleus, decreasing to 600--800\,cm$^{-3}$ at 1\arcsec\ from the nucleus and to 200--400\,cm$^{-3}$ at 2\arcsec from the nucleus. 

The [N\,II]/H$\alpha$ line ratio presents values of 1.2--2.0 within the inner 2\arcsec\ (230\,pc) and within a region extending from 2\arcsec to 4\arcsec to the SE of the nucleus, which are values typical of LINERs. The line ratio values decreases to 0.8--1.0 beyond these regions. The [O\,I]/H$\alpha$ ratio present values between 0.8--1.2 in the inner 0\farcs8 (90pc) and between 0.2--0.5 in the other regions. 

Errors in the flux distribution of the [N\,II] emission line are of the order of 5\% in the inner 1\arcsec\ and $\approx$\,10\% elsewhere. Errors in the flux distribution of H$\alpha$ are between 10--15\%. Errors in the flux distribution of the [O\,I] emission line are $\approx$\,10\% in the inner 1\arcsec\ and 20\% elsewhere. Errors in the flux distribution of the [S\,II] lines are of the order of 10\% in the inner 1\farcs5 and between 15--20\% elsewhere.

\begin{figure*}
\includegraphics[scale=1.2]{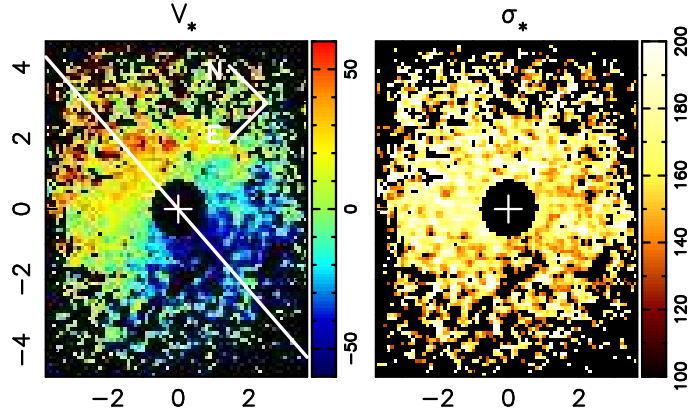}
\caption[Gaseous Kinematics]{Stellar centroid velocity field (km\,s$^{-1}$) and velocity dispersion (km\,s$^{-1}$). The straight white lines marks the position of the kinematic major axis.}
\label{figure5}
\end{figure*}

\begin{figure*}
\includegraphics[scale=1.2]{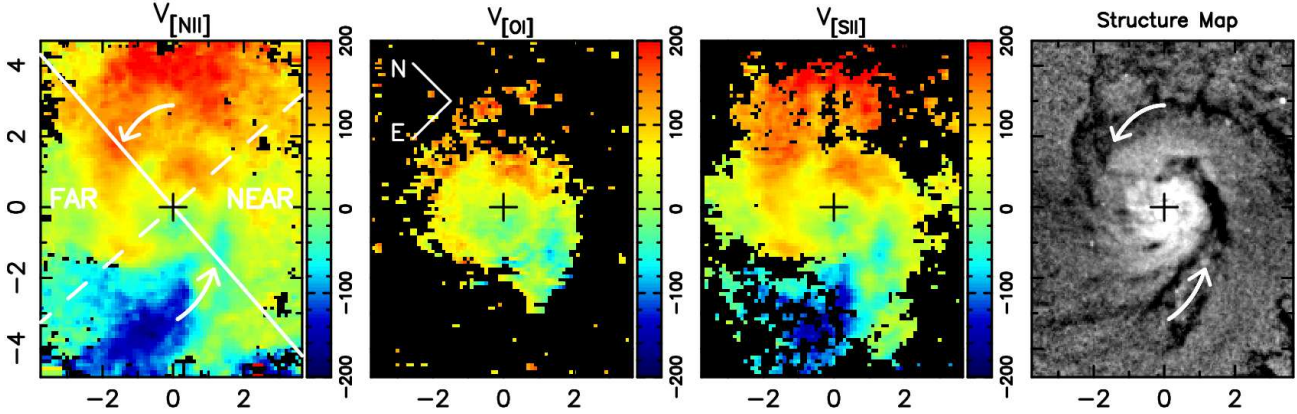}
\caption[Gaseous Kinematics]{Gaseous centroid velocities (km\,s$^{-1}$) for the [N\,II], [O\,I] and [S\,II] emission lines shown together with the structure map (rightmost panel). The straight white lines marks the position of the kinematic major axis and the dashed white lines that of the minor axis.}
\label{figure6}
\end{figure*}

\begin{figure*}
\includegraphics[scale=1.2]{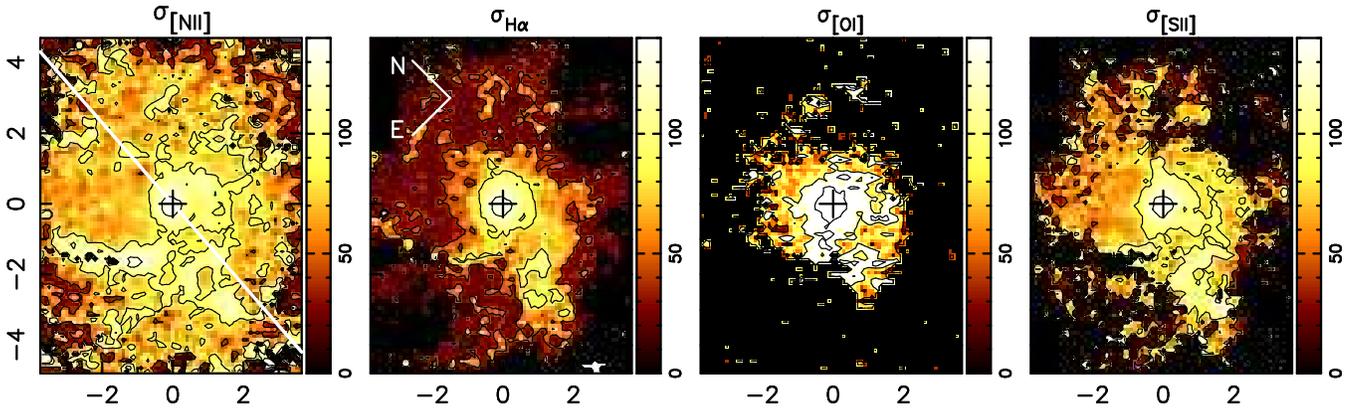}
\caption[Gaseous Kinematics]{Velocity dispersion (km\,s$^{-1}$) for the [N\,II], H$\alpha$, [O\,I] and [S\,II] emission lines. The straight white line marks the position of the kinematic major axis.}
\label{figure7}
\end{figure*}

\subsection{The stellar kinematics}\label{stellarkinematics}

The stellar velocity field V$_{*}$ is shown in the left panel of Fig.\,\ref{figure5}. It displays a rotation pattern reaching small amplitudes of $\approx$\,50\,km\,s$^{-1}$ within our field of view, with the line of nodes oriented approximately along the North--South direction, with the S side approaching and the N side receding.  The stellar velocity dispersion is shown in the right panel of Fig.\,\ref{figure5}. It varies between 140\,km\,s$^{-1}$ and 200\,km\,s$^{-1}$ across the FOV. Inspection of the nuclear spiral structure shows stronger obscuration to the West than to the East, and we thus conclude that W is the near side of the galaxy.

In the inner 0\farcs8 it was not possible to measure the stellar kinematics due to strong emission from the AGN continuum. Uncertainties in the centroid velocities and velocity dispersion are of the order of 20\,km\,s$^{-1}$. Uncertainties in the Gauss-Hermite moments \textit{{h$_{3}$}} and \textit{h$_{4}$} are of the order of 0.04, higher than the measured values, so we do not considered them in our analysis.

\subsection{The gas kinematics}\label{gaskinematics}

Centroid velocity maps for the emission lines are shown in Fig.\,\ref{figure6} along with the structure map. The gas velocity field is completely different from the stellar one. And from the comparison between the leftmost and rightmost panels of Fig.\,\ref{figure6}, it can be concluded that there is a correlation between the velocity field and the nuclear spiral seen in the structure map. Mostly blueshifts are observed to the East, the far side of the galaxy, and mostly redshifts are observed to the West, the near side, with the highest velocities following the spiral pattern seen in the structure map. Velocities of up to $\approx$200\,km\,s$^{-1}$ are observed in the [N\,II] and [S\,II] velocity maps to the NW and SE, up to $\approx$\,3\arcsec\ from the nucleus.

Maps of the H$\alpha$ and [N\,II] velocity dispersions (hereafter $\sigma_{H\alpha}$ and $\sigma_{[N\,II]}$) are shown in Fig.\,\ref{figure7}. The $\sigma_{H\alpha}$ and $\sigma_{[N\,II]}$ maps present similar structures, in spite of the fact that $\sigma_{H\alpha}$ present lower values than $\sigma_{[N\,II]}$ over the whole FOV. We attribute this difference to the underlying stellar absorption in H$\alpha$, which makes part of the H$\alpha$ emission ``fill'' this absorption resulting in an observed narrower profile. In the nucleus, both $\sigma_{[N\,II]}$ and $\sigma_{H\alpha}$ reach $\approx$\,140\,km\,s$^{-1}$, decreasing to 100\,km\,s$^{-1}$ at a radius of 0\farcs6. $\sigma_{[N\,II]}$ quickly decreases to $\approx$\,80\,km\,s$^{-1}$ to the NE and N directions while values of $\approx$\,100\,km\,s$^{-1}$ are still observed at a 2\arcsec\ radius in other directions. A similar pattern is observed in $\sigma_{H\alpha}$ and $\sigma_{[S\,II]}$.

Errors in the centroid velocity and velocity dispersion measurements for the [N\,II] emission line vary in the range 5--15\,km\,s$^{-1}$. For H$\alpha$ they vary between 5 and 20\,km\,s$^{-1}$. Errors in the [O\,I] centroid velocity and velocity dispersion are of the order of $\approx$\,10\,km\,s$^{-1}$ in the inner 1\arcsec\ and between 15 and 20\,km\,s$^{-1}$ elsewhere. For [S\,II] they are $\approx$\,5\,km\,s$^{-1}$ in the inner 1\arcsec\ and between 10 and 20\,km\,s$^{-1}$ elsewhere. 

\section{discussion}\label{Discussion}

\begin{figure*}
\includegraphics[scale=1.2]{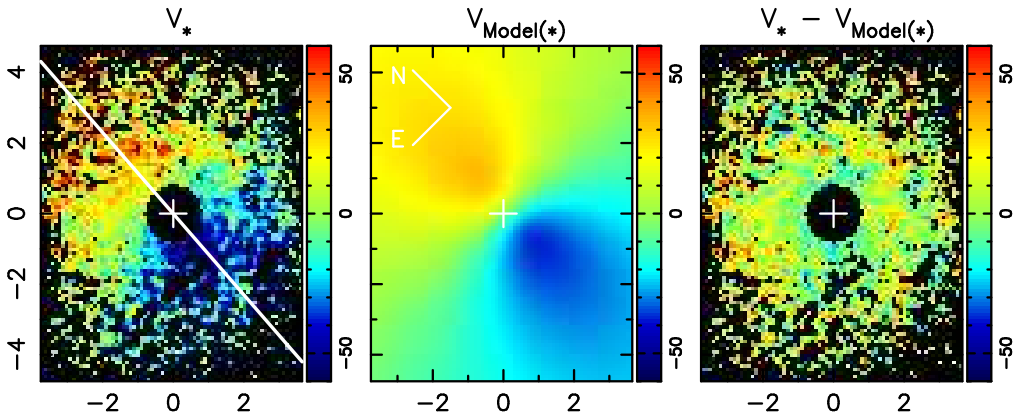}
\caption[Gaseous Kinematics]{Stellar velocity field (km\,s$^{-1}$), rotation model and the residuals between the two}
\label{figure8}
\end{figure*}

\begin{figure*}
\includegraphics[scale=1.2]{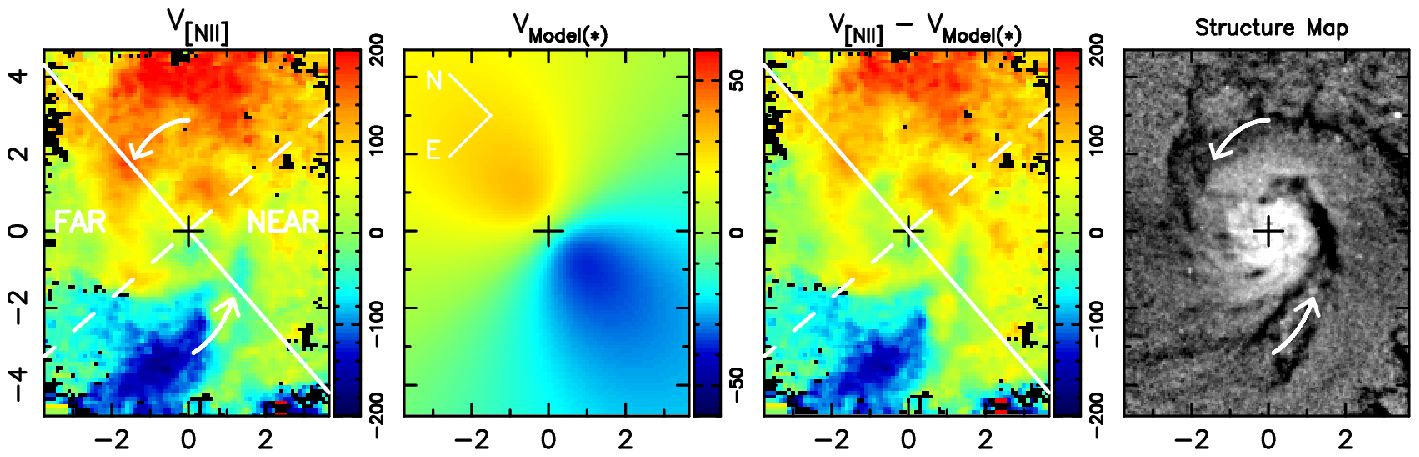}
\caption[Gaseous Kinematics]{Gaseous velocity field (km\,s$^{-1}$), stellar rotation model and residuals between the two, and the structure map.}
\label{figure9}
\end{figure*}

\subsection{The stellar kinematics}

Estimates of the position angle of the photometric major axis of NGC\,7213 in the literature range from 65\ensuremath{^\circ} to 148\ensuremath{^\circ} as the small inclination of the galaxy -- practically face on -- makes its precise determination difficult. Our data allows us to obtain the PA of the kinematic major axis, or line of nodes, via the modelling of the stellar velocity field. We have thus fitted a circular rotation model to the stellar velocity field, obtaining also the systemic velocity of the galaxy. Assuming circular orbits in a plane and a spherical potential, the observed radial velocity at a position ($R,\psi$) in the plane of the sky is given by \citep{bertola91}:
\footnotesize
\begin{displaymath}
V=V_{s}+\frac{ARcos(\psi-\psi_{0})sin(\theta)cos^{p}\theta}{\{R^{2}[sin^{2}(\psi-\psi_{0})+cos^{2}\theta cos^{2}(\psi-\psi_{0})]+c^{2}cos^{2}\theta \}^{p/2}} 
\end{displaymath}
\normalsize
where $\theta$ is the inclination of the disk (with $\theta$\,=\,0 for a face-on disk), $\psi_{0}$ is the position angle of the line of nodes, $V_{s}$ is the systemic velocity, $R$ is the radius and $A$, $c,$ and $p$ are parameters of the model. We assumed the kinematical centre to be cospatial with the peak of the continuum emission and the inclination of the disk to be 25\ensuremath{^\circ}, obtained from the apparent axial ratio (from NED\footnote{NASA/IPAC extragalactic database}) under the assumption of a thin disk geometry. The model velocity field and residuals are shown in Fig.\,\ref{figure8}.

The resulting parameters $A$, $c$, and $p$ are $133\,\pm$7\,km\,s$^{-1}$, $1\arcsec\pm0.1$ and $1.5\,\pm0.1$ respectively. The systemic velocity corrected to the heliocentric reference frame is $1648\,\pm$6\,km\,s$^{-1}$ (taking into account both errors in the measurement and the fit) and the PA of the kinematic major axis is -4\ensuremath{^\circ}\,$\pm$1. 

We can compare our derived orientation for the kinematic major axis with previous determinations from the literature. \citet{rc3} lists the photometric major axis position angle as 124\ensuremath{^\circ} and \citet{upsalla} as 45\ensuremath{^\circ}. An inspection of the H$\alpha$ image presented in \citet{hameed99} suggests an orientation of the inner gas disk of $\approx$\,40\ensuremath{^\circ}. The 2MASS catalogue \citep{2mass} lists the photometric major axis PA as 20$\ensuremath{^\circ}$, in agreement with the orientation of the kinematic major axis obtained by \citet{thaisa96} from the modelling of the large scale ionised gas velocity curve. As one can see, these different determinations are discrepant, which illustrates the difficulty in determining the orientation of the photometric major axis of NGC\,7213 due to its low inclination. Besides, the gaseous kinematics of NGC\,7213 is highly disturbed, thus a modelling of the large-scale ionised gas velocity curve does not lead to a reliable 
determination of the orientation of the kinematic major axis. Considering this, we argue that our determination of the kinematic major axis PA based on the stellar kinematics is the most reliable.

We assume the stellar velocity dispersion of the bulge to be the average stellar velocity dispersion in our FOV, corrected for the instrumental resolution of 66\,km\,s$^{-1}$, which is 177\,km\,s$^{-1}$. This value is in good agreement with previous measurements \citep{corsini03,nelson95}. Using the M-$\sigma$ relation from \citet{gultekin09}, we obtain a black hole mass of M$_{BH}$\,=\,8\,$_{-6}^{+16}\times$10$^{7}$\,M$_{\odot}$.

\subsection{The gas kinematics}
\label{gaskinexc}

As pointed out above, a comparison between the gas and stellar velocity fields (Fig.\,\ref{figure6}) shows that the two are completely distinct. The stellar velocity field has a line of nodes along PA\,=\,-4\ensuremath{^\circ}, and maximum amplitude of $\approx$\,40\,km\,s$^{-1}$, while the gas velocity field has the largest velocity gradient along PA\,$\approx$305\ensuremath{^\circ} (thus at an angle of 50\ensuremath{^\circ} with the stellar line of nodes) and much larger amplitudes, of $\approx$\,200\,km\,s$^{-1}$. Also, velocities of $\approx$\,100\,km\,s$^{-1}$ are observed in the gaseous velocity field along the stellar minor axis (PA 266\ensuremath{^\circ}), and thus it can be concluded that the gas kinematics is dominated by non-circular motions.

Our gaseous kinematics can be compared with that obtained in  previous observations. Long-slit spectroscopy along PA\,=\,50\ensuremath{^\circ} of the inner 4\,kpc of NGC\,7213 presented in \citet{thaisa96} (see Fig.\,15 in their paper) also has shown that, in the inner $\approx$1.5\,kpc, the ionised gas velocity curve is highly disturbed. Long-slit observations on larger scales -- over the inner 40\arcsec\ (4.6\,kpc) -- were obtained by \citet{corsini03} along PA's 34\ensuremath{^\circ} and 124\ensuremath{^\circ}, and show that the stellar and ionised gas velocity curves at these larger scales continue to be distinct and to have different amplitudes; they also point out that along PA\,124\ensuremath{^\circ} the gaseous velocity curve is highly disturbed up to 40\arcsec. H\,I observations \citep{hameed01} have shown that the large scale neutral gas kinematics (which extends well beyond the optical disk) is also highly disturbed, what they attribute to a 
previous merging event. 

All the available observations --  ours and the previous ones discussed above  -- thus show that the ionised and neutral gas velocity fields are disturbed in all scales. And if the gas we are observing was acquired in a previous merging event, as indicated by the H\,I data, it is not surprising that the gaseous kinematics is distinct from the stellar one. In order to try to understand the gas kinematics, a relevant question is if the gas is in the plane of the galaxy.  A clue to answer this question is the observation of the 2\,kpc  ring of  H\,II regions, which was probably formed due to the capture of gas in the merging event. This ring is almost circular and its apparent geometry follows that of the galaxy continuum image. Under the assumption that it is circular in the plane of the galaxy, \citet{corsini03} estimated an inclination of the H\,II ring of i\,=\,30\ensuremath{^\circ}, just a little larger than our estimate of  i\,=\,25\ensuremath{^\circ} (from the photometric major and minor axes diameters). 
We thus assume that the gas is indeed contained in the plane of the galaxy. 

Although the gaseous velocity field seems to be dominated by non-circular motions, if the gas is contained in the plane of the galaxy, its kinematics may include a rotation component. Under the assumption that the stellar model velocity field is a good representation of the rotation component, we illustrate, in Fig.\,\ref{figure9}, the result of the subtraction of the model from the gas velocity field. From left to right, the panels show the gas velocity field, the model and the residuals between the two, as well as the structure map, for comparison.
The largest residuals are observed along the two spiral arms (see also the structure map) to the NW and SE of the nucleus. Assuming that the gas is in the plane of the galaxy and considering that the residuals in the NW arm are observed in redshift in the near side of the galaxy, and that the residuals in the SE arm are observed in  blueshift in the far side, we conclude that we are observing inflows towards the nucleus in these two regions. The increase in the gas velocity dispersion near the borders of the spiral arms (Fig\,\ref{figure7}) can be interpreted as due to shocks in the gas as it streams towards the centre of the galaxy.

\subsection{Estimating the mass inflow rate in ionised gas}

\subsubsection{Method 1}
\label{inflow1}

The gas kinematics suggests that the ionised gas is flowing towards the nucleus. In order to estimate the mass inflow rate, we assume that both  spiral arms channel gas towards the nucleus. Assuming a similar geometry for the two arms, the mass inflow rate will be two times that along one arm. We estimate the ionised gas mass inflow rate which crosses a section of one spiral arm as:

\begin{equation}
\dot{M}_{in}\,=\,N_{e}\,v\,\pi\,r^{2}\,m_{p}\,f
\end{equation}
where $N_{e}$ is the electron density, $v$ is the velocity of the inflowing gas , $m_{p}$ is the mass of the proton, $r$ is the cross section radius of the spiral arm and $f$ is the filling factor. The filling factor can be estimated from: 
\begin{equation}
L_{H\alpha}\,\sim\,f\,N_{e}^{2}\,j_{H\alpha}(T)\,V
 \end{equation}
where $j_{H\alpha}(T)$\,=\,3.534$\,\times\,10^{-25}$\,erg\,cm$^{-3}$\,s$^{-1}$ \citep{osterbrock89} and $L_{H\alpha}$ is the H$\alpha$ luminosity emitted by a volume $V$.

Substituting Eq.\,2 into Eq.\,1 and assuming the volume of the spiral arms in the inner 1\arcsec\ can be approximated by the volume of a cone with a radius $r$ (at the base of the cone) and height $h$, we have:
 
\begin{equation}
\dot{M}_{in}\,=\frac{3\,m_{p}\,v\,L_{H\alpha}}{j_{H\alpha}(T)\,N_{e}\,h}
\end{equation}
 
We adopt as the inflow velocity that observed in the redshifted NW arm (Fig.\,\ref{figure9}) after correction for the inclination of i\,=\,25\ensuremath{^\circ}), that results in $v=130$\,km\,s$^{-1}$ in the plane of the galaxy. We consider the approximate conical region extending from the nucleus to $h\approx$\,1\arcsec\ NW, with a base radius of $r=0\farcs6$. We use the average gas density  of 472\,cm$^{-3}$ and the total H$\alpha$ flux of this region of 8.7$\,\times\,$10$^{-14}$\,erg\,cm$^{-2}$\,s$^{-1}$. Considering a distance to NGC\,7213 of 23.6\,Mpc, we obtain $L_{H\alpha}$\,=\,5.8$\,\times\,$10$^{39}$\,erg\,s$^{-1}$. For h\,=\,1\arcsec\ (115\,pc), we obtain a rate of mass inflow of $\dot{M}_{in}$\,$\approx$\,0.07\,M$_{\odot}$\,yr$^{-1}$.

\subsubsection{Method 2}

\begin{figure*}
  \centering
  \includegraphics[scale=0.5]{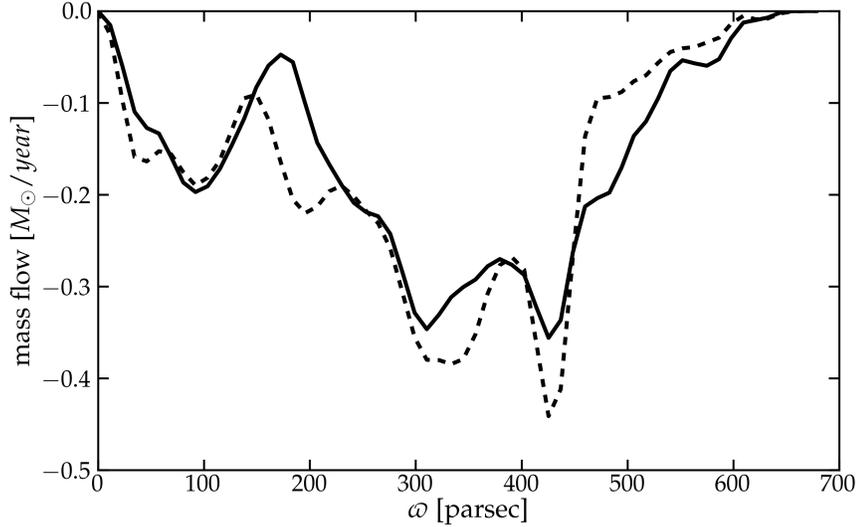}
  \caption{Mass flow rate $\dot{M}$ obtained using the residual velocity field (after subtracting the stellar rotation component -- solid curve) and from the observed velocity field (without subtracting the stellar rotation component -- dashed curve). $\varpi$ is the distance from the nucleus in the plane of the galaxy.}
  \label{figure10}
\end{figure*}

In this section we describe a new method to estimate the mass flow rate towards the nucleus: we integrate the ionised gas mass flow rate through concentric rings in the plane of the galaxy (around the whole perimeter). In order to do this, we consider that
the observed line-of-sight velocity, $v_{LOS}$ is the result of the projection of three velocity components in a cylindrical coordinate system at the galaxy. The cylindrical  coordinates are: $\varpi$ -- the radial coordinate; $\varphi$ -- the azimuthal angle;  $z$ -- the coordinate perpendicular to the plane. The galaxy inclination relative to the plane of the sky is $i$. We assume that gas vertical motions in the disc are negligible, i.e. $v_z= 0$, and then we consider two possibilities, as follows. 

\begin{itemize}

\item We first consider that the azimuthal velocity component $v_\varphi$ can be approximated by the model fitted to the stellar rotation described in section\,\ref{Discussion}, and subtract it from $v_{\rm LOS}$ in order to isolate the radial velocity component. As the deprojection of the resulting radial component into the plane of the galaxy is not well determined along the galaxy line of nodes due to divisions by zero, we have masked out from the velocity field a region of extent 0\farcs3 to each side of the line of nodes in the calculations.  

The gas mass flow rate is given by:
\begin{equation}
\label{eq:mass_flow_v.A}
\dot{M} = \rho f \  \mathbf{v}\cdot\mathbf{A},
\end{equation}
where $\rho$ is the ionised gas mass density, $f$ is the filling factor (determined via Eq.\,2), $\mathbf{v}$ is the radial velocity vector and $\mathbf{A}$ is the area vector through which the gas flows. Since we are interested in the radial flow, the area we are interested in is perpendicular to the radial direction, and thus $\mathbf{v}\cdot\mathbf{A} = v_\varpi\, A$, the product between the area crossed by the flow and the radial velocity component.

The filling factor $f$ is obtained via Eq.\,2 considering a volume $V$\,=\,$A\,d\varpi$, of a thick ring sector with width $d\varpi$ and cross-section area A, the same as above. Replacing $f$ obtained via Eq.\,2 into Eq.\,4, we finally have:
\begin{equation}
\dot{M}(\varpi) = \frac{m_p \, v_{\varpi} \, L_{\rm H\alpha}}{j_{H\alpha}(T)\, N_e \,d\varpi}.
\label{eq:dotm}
\end{equation}

Integrating Eq.\,~\ref{eq:dotm} all around the ring of radius $\varpi$, and taking into account the radial and azimuthal dependence of $v_{\varpi}$, $L_{\rm H\alpha}$ and $N_e$, we obtain the gas mass flow rate $\dot{M}$ as a function of the radius $\varpi$. It is worth noting that the mass flow rate as calculated above does not depend on $A$ and $d\varpi$, as these quantities cancel out in Eq.\,\ref{eq:dotm}, when we use $L_{\rm H\alpha}$ as the gas luminosity of a volume $V$\,=\,$A\,d\varpi$.

We have evaluated $\dot{M}(\varpi)$ through concentric Gaussian rings (rings whose radial profile is a normalised Gaussian) from the galaxy centre up to the maximum radius of 700\,pc (in the galaxy plane). The result is shown in Fig.\ref{figure10}. The negative values mean that there is inflow from the largest radius covered by our measurements down to the centre, with the mass inflow rate being largest between 400\,pc and 300\,pc, and decreasing inwards.

\item A second approach is to calculate the inflow rate without subtracting the rotational component. One might indeed question if the subtraction of the stellar velocity field from the gaseous one is adequate to isolate non-circular motions, as the stellar velocity field is subject to asymmetric drift and the gas is not. In addition, the gaseous velocity field seems to be dominated by non-circular motions. Another argument in favour of not subtracting the rotational component is to consider that its contribution to $v_{\rm LOS}$ may cancel out in the integration of the mass flow rate around the ring if the gas density and filling factor have  cylindrical symmetry, as by definition, the rotational component has cylindrical symmetry.

We have thus repeated the calculation of the mass inflow rate without subtracting the circular velocity field. The results are shown as a dashed line in Fig.\,\ref{figure10}). The difference between the mass inflow rates at any given radius between the two calculations (subtracting the rotation component and without subtracting it) is, on average, 10-15\%, showing that the resulting mass inflow rate is practically independent from the subtraction of the rotation component, supporting the assumption that the gas density and filling factor indeed approximately have cylindrical symmetry.

\end{itemize}

We can now compare the values of the mass flow rates obtained with methods 1 and 2: at a distance of 1\arcsec (115\,pc) from the nucleus, the net mass inflow rate using Method 2 is $\approx$\,0.2$\,M_{\odot}$\,yr$^{-1}$, about three times larger than the estimated $\approx$\,0.07\,M$_{\odot}$\,yr$^{-1}$ using Method 1. In Method 1 we have to rely on the observed morphology of the flow, which is not always clear, while in Method 2 we do not need to make assumptions, and just integrate the mass flow through a closed perimeter around the nucleus. For this reason, we consider Method 2 more robust than Method 1.

Finally, we point out that our calculations refer only to the ionised gas mass. If there is mass inflow in neutral and molecular gas as well, our calculated mass inflow rates can be considered lower limits to the actual mass inflow rate to the nuclear region.

We now compare the above ionised gas mass inflow rates to the accretion rate to the AGN in NGC\,7213, calculated as follows:
\[
\dot{m}\,=\,\frac{L_{bol}}{c^{2}\eta} 
\]
where $\eta$ is the efficiency of conversion of the rest mass energy of the accreted material into radiation. For LINERs it has been concluded that, in most cases, the accretion disk is geometrically thick, and optically thin \citep{nemmen06, yuan07}. This type of accretion flow is known as RIAF (Radiatively Inefficient Accretion Flow \citep{narayan05}), and has a typical value for $\eta$ of $0.01$ ($1\%$), although it can be as low as $0.001$ ($0.1\%$) \citep{xie12}. The bolometric luminosity of the active nucleus was estimated by \citet{emmanoulopoulos12} as $L_{bol}$\,=\,1.7\,$\times\,$10$^{43}$\,erg\,s$^{-1}$. We use these values to derive an accretion rate of $\dot{m}$\,=\,3$\,\times\,$10$^{-2}$\,M$_{\odot}$\,yr$^{-1}$ for an efficiency of $1\%$ and $\dot{m}$\,=\,3$\,\times\,$10$^{-1}$\,M$_{\odot}$\,yr$^{-1}$ for an efficiency of 0.1$\%$. The ionised gas mass inflow rate at 1\arcsec\ from the nucleus, as estimated from Method 2, ranges between seven times larger to the approximate value of the 
accretion rate.   

Our observations may imply that the accretion rate to the AGN will increase in the future. Nevertheless, gas at these distances has to lose 99.99$\%$ of its angular momentum before it reaches the central BH \citep{jogee06}. This may result in the accumulation of gas in the circumnuclear region and the subsequent triggering of star formation. Recent simulations support this scenario \citep{hopkins10}. Previous integral field observations by our group and others, have indeed revealed the presence of nuclear rings (at hundreds of parsecs from the nucleus) with stellar population dominated by young to intermediate-age stars in a number of active galaxies \citep{davies07,barbosa06,rogemar10a,riffel11,thaisa1068}. 

These rings suggest the association of the formation of stars tens to hundred million years ago with the onset of the nuclear activity, favouring the evolutionary scenario proposed by \citet{thaisa01} and/or that of \citet{davies07}. At an inflow rate of 0.2\,M$_{\odot}$\,yr$^{-1}$, a reservoir of $\approx$\,2$\,\times\,$10$^{6}$\,M$_{\odot}$ in ionised gas alone can be built up in just 10$^7$yr, and can fuel the formation of new stars in the bulge.

\section{conclusions}\label{Conclusion}

We have measured the gaseous kinematics in the inner 0.8\,$\times$\,1.1\,kpc$^2$ of the LINER/Seyfert\,1 galaxy NGC\,7213, from optical spectra obtained with the GMOS integral field spectrograph on the Gemini South telescope at a spatial resolution of $\approx$\,60\,pc. The main results of this paper are:

\begin{itemize}

\item The stellar velocity field shows high velocity dispersions of up to 200\,km\,s$^{-1}$ and circular rotation with an orientation for the line of nodes of $\approx$\,-4$^\circ$ ($\approx$\,N). On the basis of the dust distribution, it can be concluded that the near side of the galaxy is the W and the far side is the E;

\item The stellar velocity dispersion of the bulge is 177\,km\,s$^{-1}$, leading to a black hole mass of M$_{BH}$\,=\,8\,$_{-6}^{+16}\times$10$^{7}$\,M$_{\odot}$;

\item The gaseous velocity field is completely distinct from the stellar one, being dominated by non-circular motions. Velocities of up to $\approx$\,200\,km\,s$^{-1}$ are observed in two spiral structures extending from the nucleus to $\approx$\,4\arcsec\ (460\,pc) NW and SE, which are correlated with spiral arms seen in a structure map of a continuum image of the region; 

\item Redshifts are observed along the NW spiral arm -- on the near side of the galaxy, while blueshifts  are observed along the SE spiral arm -- on the far side; as the observations also suggest that the gas is in the galaxy plane, we conclude that the gas is inflowing towards the centre along the spiral arms. An observed increase in the gas velocity dispersion, cospatial with the nuclear spiral -- interpreted as shocks in the inflowing gas --  supports this conclusion; 

\item We estimate the ionised gas mass flow rate towards the nucleus using two methods. In the first method, we use an approximate geometry for the flow along the spiral arms and obtain a mass flow rate of $\approx$\,0.07\,M$_{\odot}$\,year$^{-1}$ at a distance of 1\arcsec\ (115\,pc) from the nucleus. In the second method we calculate the net gas mass flow rate across a series of concentric rings around the nucleus, obtaining a mass flow rate ranging from 0.4\,M$_{\odot}$\,year$^{-1}$ at 400\,pc from the nucleus down to 0.2\,M$_{\odot}$\,year$^{-1}$ at 100\,pc. We conclude that the second method is more robust as it does not depend on the exact geometry of the flow, which is not that clear from the data;

\item As our observations are only of the ionised emitting gas, and the inflow should include also neutral and molecular gas, the inflow rates quoted above are actually lower limits of a probably much larger gas mass inflow rate;

\item Considering that gas at 100 pc scales needs to lose more than 99\% of its angular momentum to reach the accretion disk, most of the gas will probably accumulate in the circumnuclear region, where episodes of star formation may occur. This is supported by previous studies in which circumnuclear rings of young to intermediate age stars are observed around AGN, and may be the process leading to the so-called co-evolution of the galaxy and its SMBH.

\end{itemize}
 
\section*{ACKNOWLEDGEMENTS}

We acknowledge the referee for relevant suggestions which have improved the paper. This work is based on observations obtained at the Gemini Observatory, which is operated by the Association of Universities for Research in Astronomy, Inc., under a cooperative agreement with the NSF on behalf of the Gemini partnership: the National Science Foundation (United States), the Science and Technology Facilities Council (United Kingdom), the National Research Council (Canada), CONICYT (Chile), the Australian Research Council (Australia), Minist\'erio da Ci\^encia e Tecnologia (Brazil) and south-eastCYT (Argentina). NN acknowledges funding from ALMA-Conicyt 31110016, BASAL PFB-06/2007, Anillo ACT1101 and the FONDAP Center for Astrophysics. This material is based upon work supported in part by the Brazilian institution CNPq. 

\bibliographystyle{mn2e.bst}
\bibliography{ngc7213.bib}

\label{lastpage}
\end{document}